\input harvmac
\input epsf
\epsfverbosetrue

\def\p{\partial}
\def\ap{\alpha'}

\def\b0{\bar{0}}
\def\b4{\bar{4}}
\Title{EFI-98-12}{\vbox{\centerline{'t Hooft Vortices on D-branes}
\vskip12pt
}}
\vskip20pt
\centerline{Miao Li}
\bigskip
\centerline{\it Enrico Fermi Institute}
\centerline{\it University of Chicago}
\centerline{\it 5640 Ellis Avenue, Chicago, IL 60637, USA}

\bigskip
\centerline{\it }
\centerline{\it }
\centerline{\it }
\bigskip
The point where a D2-brane intersecting a stack of D2-branes is 
proposed as a candidate for the 't Hooft vortex in the world-volume
theory of N D2-branes. This straightforwardly generalizes to 
D3-branes, where a vortex line is generated by the intersection.
Similarly, there are such objects on M-branes. We use Maldacena's
conjecture to compute the static potential between a vortex and
an anti-vortex in each case, in the large N limit.

\Date{March 1998}

\nref\th{G. 't Hooft, Nucl. Phys. B138 (1978) 1.}
\nref\tamiaki{T. Yoneya, Nucl. Phys. B144 (1978) 195.}
\nref\cm{C. Callan and J. Maldacena, hep-th/9708147.}
\nref\ewm{E. Witten, hep-th/9703166.}
\nref\jm{J. Maldacena, hep-th/9711200.}
\nref\gkp{S. Gubser, I. Klebanov and A. Polyakov, hep-th/9802109.}
\nref\ewde{E. Witten, hep-th/9802150. }
\nref\coll{I. R. Klebanov, hep-th/9702076; S. Gubser, I. R. Klebanov
and A. A. Tseytlin, hep-th/9703040; M. R. Douglas, J. Polchinski
and A. Strominger, hep-th/9703031; A. Polyakov, hep-th/9711002.}
\nref\jmald{J. Maldacena, hep-th/9803002.}
\nref\ry{S. Rey and J. Yee, hep-th/9803001.}
\nref\minahan{J. Minahan, hep-th/9803111.}
\nref\witten{E. Witten,  hep-th/9803131.}
\nref\rty{S. Rey, S. Theisen and J. Yee, hep-th/9803135.}
\nref\bisy{A. Brandhuber, N. Itzhaki, J. Sonnenschein and 
S. Yankielowicz, hep-th/9803137.}
\nref\imsy{N. Itzhaki, J. Maldacena, J. Sonnenschein and
S. Yankielowicz, hep-th/9802042.}

\newsec{Introduction} 

't Hooft in \th\ introduced in a nonabelian gauge theory a new 
variable dual to the Wilson loop. If the Wilson loop is regarded
as the usual order parameter, the dual variable may be regarded
as a disorder parameter. Thus the dual variable is also useful
in characterizing the phases of the nonabelian gauge theory.

There are two possible interpretations of
the usual Wilson loop, depending on whether it is temporal or
it is spatial. In the temporal case, the loop may be explained as
traced out by a charge. For instance, if the charge is in the
fundamental representation $R$ of the gauge group $SU(N)$, then
the corresponding Wilson loop is just 
\eqn\wloop{W(C)=\tr_R P e^{i\int_C A},}
where the trajectory $C$ is parametrized by time.
If the loop is spatial, the operator may be taken as a creation 
operator for a string-like object. Indeed there is an electric flux
along the loop in this case.

The disorder parameter is a scalar field in $2+1$ dimensions, and
a loop variable in $3+1$ dimensions. In the first case, there is
an exchange relation
\eqn\exch{W(C)\phi (x)=e^{{2\pi i\over N}}\phi (x)W(C),}
where again we assumed gauge group $SU(N)$. Note that loop $C$ 
surrounds point $x$ exactly once. If $C$ and $x$ are not ``linked",
there is no extra phase in the exchange relation. The effect of 
operator $\phi (x)$ is to introduce a gauge transformation which
is singular at point $x$. Namely, if one follows a loop surrounding
$x$, the gauge transformation is multi-valued: $G\rightarrow
e^{2\pi i/N}G$. The phase factor belongs to the center $Z_N$ of
group $SU(N)$, therefore introduces no effects on fields in the
adjoint representation.

The analogue of $\phi(x)$ in $3+1$ dimensions is a loop variable.
For a loop $C$, denote this operator by $T(C)$. If $C$ and $C'$
have link number 1, then \exch\ generalizes to
\eqn\link{T(C)W(C')=e^{{2\pi i\over N}}W(C')T(C).}
We call $T(C)$ the 't Hooft loop operator.

One might speculate that the 't Hooft loop in a $3+1$ dimensional
gauge theory ought to correspond to a loop traced out by a monopole.
There are doubts whether this can be true. Indeed, a monopole is
nonlocal with respect to an electric charge. It is so only when
the monopole and the electric charge live in the same $U(1)$ group.
Now for $SU(N)$, there are $N-1$ ``fundamental'' monopoles.
If we were to construct a loop variable associated to monopoles,
we must include all of them, just as in the case we construct
the Wilson loop. Since different electric charges and different
monopoles are not all nonlocal with respect to one another, it
is hard to imagine the exchange relation \link\ will result 
from this construction. Another problem associated with this is,
if a trace of monopole is the 't Hooft loop, what object generates
$\phi(x)$ in a $2+1$ gauge theory? Similar to the case of the 
spatial Wilson loop, the spatial 't Hooft loop on D3-branes then 
would have to be interpreted as an operator creating a trapped
D-string. By T-duality, a vortex on D2-branes would have to be
interpreted as a trapped D0-brane. It is well-known that a 
D0-brane, once trapped in D2-branes, can not be localized. We
conclude that a D0-brane can not be a candidate for a vortex,
hence the loop traced by a monopole can not be a candidate for
a 't Hooft loop.

Our construction depends crucially on the existence of the adjoint scalars 
in the super Yang-Mills theory.
It is possible to construct the 't Hooft loop operator in a gauge 
theory without adjoint Higgs fields, as done in \tamiaki. 

In the next section, we propose that there is a natural realization
of 't Hooft vortex in the world-volume theory of D2-branes, it is
just given by the touch point on which an orthogonal D2-brane 
intersects the stack of D2-branes in question. Our construction is
a simple generalization of an observation made in \cm. We argue that
precisely for an operator which acts as the creation operator for
this touch point, the exchange relation \exch\ holds. The 
generalization of this to a stack of D3-branes is obvious: The 't
Hooft loop is just the loop along which an orthogonal D3-brane
intersects the stack of D3-branes. In sect.3, we use Maldacena's
conjecture to compute the static potential between a vortex and
an anti-vortex, in the large
N limit. This calculation is generalized to D3-branes in sect.4,
and to M-branes in sect.5.

\newsec{'t Hooft vortex and intersecting D-branes}

The simple example we start with is two intersecting D2-branes, each
lying on a different complex plane. Consider the four dimensional
subspace $X^a=0, \quad a=5, \dots , 9$. This subspace can be described
by two complex variables $z=X^1+iX^2$, $w=X^3+iX^4$. Take a D2-brane
lying along the $z$ plane at $w=0$, and another D2-brane lying along
the $w$ plane at $z=0$. The SUSY preserved by the first D2-brane
is one subject to the constraint $\epsilon =\gamma^0\gamma^{12}
\tilde{\epsilon}$, while the SUSY preserved by the second D2-brane
is the one satisfiying $\epsilon =\gamma^0\gamma^{34}\tilde{\epsilon}$.
It is easy to check that the two conditions are compatible, and
only $1/4$ of the whole SUSY is preserved by this configuration.

The above two D2-branes intersect at the point $(z,w)=(0,0)$. The
surface can be described by the simple equation $zw=0$. It turns
out that there is a family of deformations of this configuration,
parametrized by one complex parameter $C$, $zw=C$. This was observed
in \cm. For $C\ne 0$, the surface is smooth at either $z=0$ or
$w=0$, but the point $(z,w)=(0,0)$ is no longer on the surface.
The parameter $C$ can be interpreted as the condensate of strings 
stretched between the two orthogonal D-branes.

The equation $zw=C$ has a natural interpretation in the world-volume
theory of one of the D-branes. If we start with the first D-brane,
then $w$ can be regarded as a complex scalar field living on the
D2-brane. The Nambu-Goto action for a static field configuration reads
\eqn\ngac{S=-T_2\int dtd^2z\left(1+(\p w\bar{\p}\bar{w}-\bar{\p} w\p 
\bar{w})^2+2(\p w \bar{\p}\bar{w}+\bar{\p}w\p\bar{w})\right)^{1/2},}
where $T_2$ is the D2-brane tension. The static energy is the minus
the above integral expression with the time integral dropped out.
The equation of motion derived from the above action is the one 
determining a minimal surface. It is easy to see that it admits 
holomorphic solutions, or anti-holomorphic solutions. For holomorphic
solutions, the energy formula simplifies to
\eqn\holoac{E=T_2\int d^2z(1+\p w\bar{\p}\bar{w}),}
and for consistency, it is readily checked that the a holomorphic
function $w(z)$ satisfies the equation of motion derived from the
simplified action.

Now substitute the equation $w=C/z$ into \holoac, 
\eqn\twdtw{E=T_2\left(\int d^2z +\int d^2z {|C|^2\over |z|^4}\right).}
The first integral gives the energy of the original D2-brane. The 
second integral diverges at $z=0$. Introduce a cut-off $r$, and
demand $|z|\ge r$, the second integral results in the additional
energy $T_2\pi (|C|^2/r^2)$. Now $R=|C|/r$ is naturally interpreted
as the infrared cut-off for the second D2-brane lying along the
complex plane $w$. The additional energy is precisely that of the
second D2-brane. Physically, we can trust the Nambu-Born-Infeld
action if the derivatives of scalar fields are not very large,
this would introduce a natural short distance cut-off $r$ on the
$z$ plane. From the above simple calculation, it appears that 
this cut-off can be arbitrarily small.

The interpretation of intersecting branes as a single, holomorphically
embedded brane was first considered in the M theory context
in \ewm. The situation we are discussing here is simpler, since
the target space is not only flat, but also has a trivial topology.

It is straightforward to generalize the above consideration to the case
when a single D2-brane along the $w$ plane intersects several
D2-branes along the $z$ plane. Assume these D2-branes intersect
the orthogonal D2-brane at points $w_i$, $i=1,\dots , N$, the general
holomorphic representation is
\eqn\genholo{z=\sum_i {C_i\over w-w_i},}
where $z$ can be interpreted as a complex scalar field on the single
D2-brane. For distinct $w_i$, the complex parameters $C_i$ are 
arbitrary. Now interesting things happen when we push all the $N$ 
D2-brane to the same point, say $w_i=0$. It is no longer possible to
keep $C_i$ arbitrary. For instance, if $C=\sum_i C_i\ne 0$, we return
to the simple case where there is only one D2-brane along the $z$
plane. $C_i$ must be fine tuned such that as a function of $w$,
$z$ has a pole of order $N$ at $w=0$. This is easy to see for $N=2$,
and for general $N$ the result can be deduced inductively. As a 
direct check of this claim, consider the energy of the configuration
$zw^N=C$. Applying formula \holoac\ with the roles of $z$ and $w$
interchanged, we find
\eqn\multie{E=T_2\left(\int d^2w +\int d^2w {N^2|C|^2\over 
|w|^{2N+2}}\right).}
Again introduce a cut-off $|w|\ge r$, the second integral
results in $NT_2\pi (|C|^2/r^{2N})$. Now $R=|C|/r^N$ is the
size cut-off for the $N$ D2-branes, and the excessive energy is 
precisely that of these $N$ D2-branes.

The Higgs field $z=C/w^N$ is well-defined on the single, orthogonal
D2-brane. The physical interpretation of this solution is rather
simple. If one follows a loop on the $w$ plane around $w=0$, one
goes $N$ loops on the $z$ plane. That is, the coincident $N$ D2-branes
are connected from one sheet to another. 
It is interesting to start with these $N$ D2-branes,
and interpret $w$ as a Higgs field. $w$ is not well-defined as 
a function of $z$, because of the branch cut. However, we know that
the world-volume theory is a nonabelian theory with gauge group
$U(N)$, and a Higgs field must be a $N\times N$ matrix. Thanks
to the branch cut, a diagonal matrix is easy to construct using
$zw^N=C$:
\eqn\nonah{W(z)=({C\over z})^{{1\over N}}\hbox{diag}
\left(1, e^{-{2\pi i\over N}},\dots , e^{-{2\pi i(N-1)\over N}}
\right).}
Still, this Higgs field configuration is not well-defined. This
is because if one follows a loop around $z=0$, the first diagonal
element will be shifted to the second, and so on. Namely
$W_a(ze^{2\pi i})=W_{a+1}(z)$. 

It is well-known how to resolve this problem. Perform a gauge
transformation 
\eqn\gauget{\tilde{W}(z)=e^{-iA\theta}W(z)e^{iA\theta},}
where $\theta$ is the angular variable of the $z$ plane, and $A$ is
a constant matrix satisfying
\eqn\will{e^{2\pi iA}=U,\quad  U_{ij}=\delta_{i,j-1}.}
Thus the gauge transformation is singular, since its value after
a full rotation is the shift matrix $U$, not $1$. Now, the
new Higgs field $\tilde{W}$ is single-valued. In addition
to this Higgs field, there is a gauge field $A_\theta =A$.
This gauge field generates a Wilson line around $z=0$, whose
effect is precisely to connect the a-th sheet to the (a+1)-th
sheet.

The solution ($\tilde{W}(z)$, $A$) can be regarded as a vortex
solution, although its energy diverges without cut-off. We propose
that this vortex is a 't Hooft vortex, namely if a creation operator
$\phi(z)$ is defined for the vortex, it will have the commutation 
relation with a Wilson loop operator as in \exch. It is quite 
difficult to prove this directly. Here we only provide two 
intuitive arguments.

The first argument has to do with the nature of the charge carried by
a vortex. Topologically, the 't Hooft vortex is possible due to the
fact that $\Pi_1(SU(N)/Z_N)=Z_N$, that is, the gauge transformation
induced by the presence of the vortex is defined up to a factor in
$Z_N$ when a loop around the vortex is followed. If there are $N$
such vortices inside the loop, the group factor becomes trivial,
and one can not distinguish between this configuration and a 
topologically trivial configuration. Thus, the charge carried by a 
vortex is different from a standard $U(1)$ charge. To see that this is
also true of the vortex we have constructed, we need the equation
describing the N D2-branes along the $w$ plane intersecting the
other N coincident D2-branes along the $z$ plane. If the intersecting
points are $z_i$, the holomorphic relation is
\eqn\znc{w^N\prod_{i=1}^N(z-z_i)=C,}
which generalizes $zw^N=C$. When the N vortices sit at the same 
position, say $z_i=0$, the above relation becomes $(zw)^N=C$.
It is obvious that the different sheets on the $z$ plane are no
longer connected. We conclude that our vortex indeed carries
a $Z_N$ charge.

The second argument is indirect evidence for the exchange relation
\exch. Consider the correlation function $\langle W(C)\phi(x)
\rangle$, where $C$ is a time-like straight line in spacetime, and
$x$ is a spacetime point whose time component is $X^0=0$. 
$W(C)$ is a temporal Wilson line, the $\phi(x)$ creates a vortex
at time $X^0=0$. If the distance between $x$ and $C$ is large 
enough, we expect the declustering behavior, namely
\eqn\dec{\langle W(C)\phi(x) \rangle=\langle W(C)\rangle \langle 
\phi(x)\rangle e^{i\alpha(C,x)},}
where the phase $\alpha$ might not be globally defined.

Now, adiabatically rotate $C$ along a loop surrounding $x$, the
phase $\alpha$ may jump by an amount $\Delta\alpha$. Physically,
the Wilson loop is a trajectory of a quark in the fundamental
representation of $SU(N)$. For a charge moving, say, on the first
brane along $C$, it will jump to the second brane at time 
$X^0=0$, after the adiabatic rotation, since the effect of the
vortex is precisely to cause the connection to the second sheet
at this time. Similarly, a charge moving on the second brane
will jump to the third after the rotation, and so on. Thus in
general there is no reason that the correlation function \dec\
should come back to its original value, thus the jump $\Delta
\alpha$ may not be zero. However, if one executes the rotation
N times, the value of the correlation function must be resumed. That
is, the condition on $\Delta\alpha$ is $N\Delta \alpha=2\pi l$
with an integer $l$. The minimal nonvanishing value is
$\Delta\alpha =2\pi/N$. This is precisely the phase appearing in
the exchange relation \exch. Here we must emphasize that our
argument is only a plausible one. We do not have a proof that
the exchange relation \exch\ is valid.
\bigskip
{\vbox{{\epsfxsize=2in
        \nobreak
    \centerline{\epsfbox{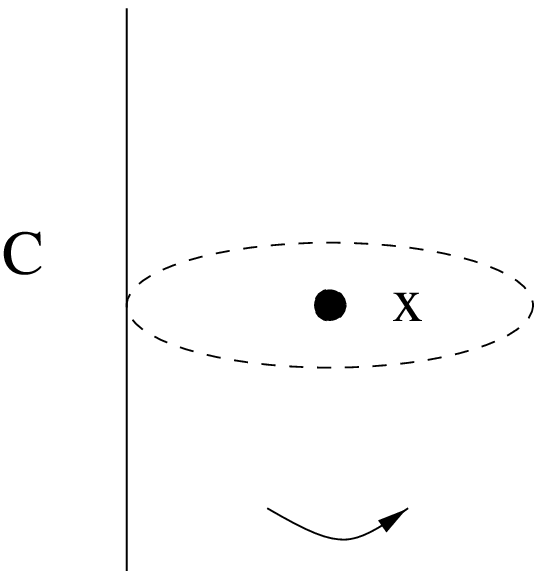}}
        \nobreak\bigskip
    {\raggedright\it \vbox{
{\bf Figure 1.}
{\it A Wilson loop is adiabatically brought about along a loop surrounding 
pint $x$.}
 }}}}}
\bigskip
't Hooft discussed correlation function such as $\langle
\phi(T)\phi^+(0)\rangle$ in \th. The physical process involved in 
this correlation is a creation of a vortex at time $X^0=0$
which subsequently propagates to time $X^0=T$, and is then destroyed.
Our vortex has an infinite mass, thus we expect that such a correlation
vanishes, since the propagator vanishes in the large mass limit.
This will be checked in a large N calculation in the next
section.

The discussion in this section has a straightforward generalization
to D-branes with higher spatial dimensions. The case of special
interest is D3-branes. Here one merely adds a common spatial
dimension to the intersecting D-branes. The intersection
appears as vortex line. This is a candidate for a spatial
't Hooft loop. Unlike the Wilson loop, we do not know how to
write down a corresponding operator for it. Also, a spatial
Wilson loop creates a state of finite mass, and it can be 
interpreted as trapped fundamental string within D3-branes.
Our vortex line is rather similar to a temporal Wilson loop.
A temporal Wilson can be regarded as traced out by a quark of
infinite mass. Here, the intersection is $1+1$ dimensional,
representing a spatial vortex line propagating in time.
An instantaneous spatial vortex line can be constructed by a 
moving D3-brane which intersects N orthogonal D3-branes only 
at a given time.

\newsec{Interaction potential between a vortex and an anti-vortex
in large N limit}
To see that what we have proposed is sensible, it is useful to calculate
some correlation functions or interaction potential. The latter
for a vortex and an anti-vortex must be necessarily negative, if
our interpretation of the configurations is meaningful. Until
the recent bold conjecture of Maldacena \jm, no tool for computing
correlation functions in a strongly coupled gauge theory had been 
available. The proposed duality between a large N conformally
invariant field theory and string/supergravity on an anti-de Sitter
space has been precisely formulated in \refs{\gkp, \ewde},
and checked to some extent. Some seeds for this conjecture appeared
in earlier work \coll. There are a large number of papers
written in a short period, for a partial list see the reference lists
in papers cited here. We will mention only those papers which are
directly related to the problem studied here.

The formalism for computing the expectation value of a Wilson loop
was proposed in \jmald\ (see also \ry ). It has been applied to
various large N gauge theories at zero temperature \refs{\jmald,
\ry, \minahan}, as well as theories at a finite temperature \refs{\witten,
\rty, \bisy}. The results are perfectly sensible physically. This
certainly lends much support to Maldacena's conjecture.

The large N gauge theory on D2-branes is not conformally invariant.
Even so, a similar duality can be valid for a certain range of 
energies, as analyzed in \imsy. Indeed, if a finite temperature
for a conformally invariant theory is switched on, the conformal
invariance is lost, nevertheless the correspondence between the
field theory and supergravity in the bulk should still be valid.
Thus, there is no reason to exclude a nonconformally invariant
theory from the conjecture.

Now, if the number of D2-branes is large enough, the gravitational
background induced by these branes can no longer be ignored.
Instead of analyzing the Nambu-Born-Infeld action in a flat 
background, as we have done in the previous section, we need to work
with the action in a curved background. In the large N limit,
a D2-brane intersecting with a large number of D2-branes does
not have an interpretation as a smooth Higgs excitation. Indeed,
the equation $zw^N=C$ does not have a well-defined large N limit.
We make a change to the complex parameter such that the new
equation is
$$z=\left({C\over w}\right)^N.$$
If we hold both $C$ and $z$ fixed and take the large N limit, we get
either $z=0$ or $z=\infty$, depending on whether $|w/C|$ is 
greater than or less than 1. Apparently the physically sensible
choice is $z=0$. In this case, if $C^N\sim l_s^{N+1}$, where $
l_s$ is the string scale, then the condition on $|w|$ is just
$|w|>l_s$.
In this case the test D2-brane intersects the
stack of D2-brane precisely at $z=0$. We shall see that this is 
compatible with the Nambu-Born-Infeld action in the curved 
background.

In the large N limit, the metric and the dilaton generated by a stack 
of coincident D2-branes in the near horizon region are
\eqn\back{\eqalign{ds^2&=\ap \left({U^{5/2}\over R^2}ds^2_3 +{R^2
\over U^{5/2}}ds^2_7\right), \cr
e^\phi &= {g_{YM}^2R\over U^{5/4}},}}
where $ds^2_3$ is the flat Minkowski metric on D2-branes, $ds^2_7$
is the flat Euclidean metric on $R^7$, and $U$ is the radial 
coordinate on $R^7$, $R^2=(6\pi^2g_{YM}^2 N)^{1/2}$. The definition
of the Yang-Mills coupling is $g_{YM}^2=g(\ap)^{-1/2}$, where $g$
is the string coupling constant. Since in all of our calculations,
factors containing $\ap$ cancel, we set $\ap =1$.

The Nambu-Born-Infeld action for a test D2-brane in the above 
background is
\eqn\curvbi{S=-{1\over 4\pi^2}\int d^3x e^{-\phi}[\det 
(G_{\mu\nu}\p_\alpha X^\mu\p_\beta X^\nu )]^{{1\over 2}}.}
For a static orthogonal D2-brane, due to the rotational invariance on 
$R^7$, we can always choose its spatial coordinates to coincide 
with two of $R^7$. Other coordinates of $R^7$ vanish, in order
for the D2-brane to intersect the source D2-branes. Denote, as in the 
previous section, the spatial
coordinates by a complex variable $w$. It is easy to see that
the factors depending on $U$ inherited from the metric and dilaton 
in \back\ all cancel, and the action takes the same form as in a flat
space, thus the energy formula for an orthogonal D2-brane agrees 
with that in a flat space. This must be the case, given the BPS
nature of this state. On the other hand, if the D2-brane also bends
in the longitudinal direction which we denote by a complex 
coordinate $z$, the action is quite different, and is given by
\eqn\bentac{S=-{1\over 4\pi^2g_{YM}^2}\int d^3x\left(1+
2{|w|^5\over R^4}(|\p z|^2+|\bar{\p}z|^2)+
{|w|^{10}\over R^8}(|\p z|^2-|\bar{\p}z|^2)^2\right)^{{1\over 2}},}
where we have identified $U$ with $|w|$, and the derivatives
are defined against the complex coordinates $w$, $\bar{w}$.
It is obvious that the equations of motion derived from this 
action do not admit holomorphic solutions, and $z=0$ minimizes 
the energy. This agrees with our observation based on solution 
in the flat background.

One can consider a solution of multiple parallel branes, again the
solution is trivial $z=z_i$, $z_i$ is the location of the i-th 
D2-brane. If, however, there is an anti-D2-brane in addition to
a D2-brane, the solution is no longer constant $z$. To minimize 
the energy cost, there will develop a throat between the D2-brane 
and the anti-D2-brane, as in fig.2. We will compute the energy 
difference between this configuration and the sum of energies of 
individual branes. The difference is finite but negative, confirming
our expectation. In analogy to the case of the Wilson loops,
we interpret the energy difference as the static potential between
a vortex and an anti-vortex in the world-volume theory in the
large N limit.

Assume the location of the D2-brane is $z=L/2$, and the location 
of the anti-D2-brane is $z=-L/2$. From the action \bentac\ which
is valid locally on the $w$ plane, we see that in order to minimize
the energy, the solution will always have $\Im z=0$. By inspecting
fig.2, it is easy to see that a good coordinate system is
provided by $(x,\theta)=(\Re z, \theta)$, where $\theta$ is
the angular variable on the $w$ plane. Now, $r=|w|$ is an even
function of $x$, due to the reflection symmetry.
The appropriate boundary conditions are $r(x=\pm L/2)=\infty$.
\bigskip
{\vbox{{\epsfxsize=2in
        \nobreak
    \centerline{\epsfbox{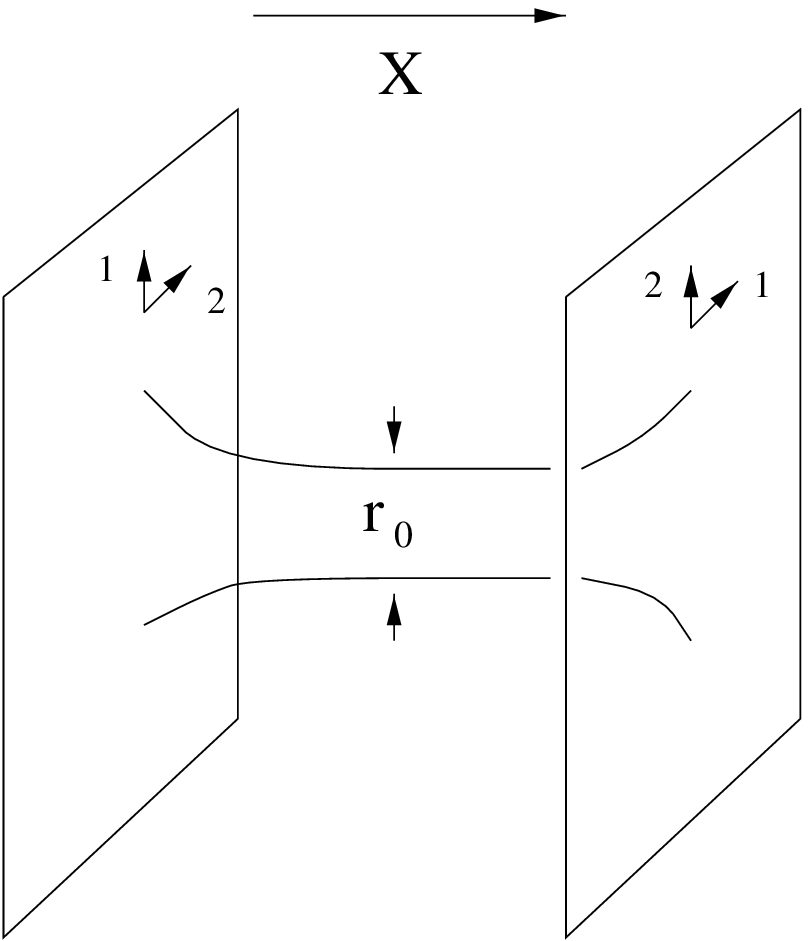}}
        \nobreak\bigskip
    {\raggedright\it \vbox{
{\bf Figure 2.}
{\it A pair of D2-brane and anti-D2-brane develop a throat between them
in the curved background to minimize the energy cost.}
 }}}}}
\bigskip

A simple calculation leads to the energy formula
\eqn\energy{E={1\over 4\pi^2g_{YM}^2R^2}\int d\theta dx
r^{{7\over 2}}\left(1+R^4r^{-5}(r')^2\right)^{{1\over 2}},}
where $r'=dr/dx$. Due to translational invariance in $x$,
the equation of motion has an integration
\eqn\integ{r^{{7\over 2}}(1+R^4r^{-5}(r')^2)^{-{1\over 2}}
=r_0^{{7\over 2}},}
where $r_0$ is the minimal value of $r$ that is reached at 
$x=0$, it can be interpreted as the size of the throat.
$r_0$ is determined by the boundary condition
\eqn\dint{R^2\int_{r_0}^\infty dr r^{-{5\over 2}}
\left(({r\over r_0})^7-1\right)^{-{1\over 2}}={L\over 2}.}
The result is
\eqn\rzero{r_0^{{3\over 2}}={4\Gamma (5/7)\Gamma(1/2) R^2
\over 3\Gamma (3/14)L}.}
This is quite reasonable, since the size of throat $r_0$ 
decreases when $L$ increases, and eventually disappears in
the limit $L=\infty$.

The static energy \energy\ is divergent, and we need to subtract
the bare energy of individual D-branes, which is
$$2\times {1\over 4\pi^2g_{YM}^2}\int_0^\infty rdrd\theta.$$
After the subtraction, the interaction potential is 
\eqn\poten{V=E-E_0=-{r_0^2\over \pi g_{YM}^2}
\left({1\over 2}-\int_1^\infty xdx[x^{{7\over 2}}
(x^7-1)^{-{1\over 2}}-1]\right).}
The number in the parenthesis is a pure number, and is positive.
According to \rzero, $r_0^2\sim (R^2/L)^{4/3}\sim (g_{YM}
\sqrt{N}/L)^{4/3}$. Thus the potential scales as 
$N^{2/3}g_{YM}^{-2/3}L^{-4/3}$, to be contrast to the potential
between a pair of heavy quark and anti-quark, which scales
as $(g_{YM}^2N)^{1/3}L^{-2/3}$ \jmald. The potential between a vortex
and anti-vortex falls off with $L$ faster than that between
quark and anti-quark, but grows faster with the large N.

The integral in \poten\ can be calculated. For instance it is
given by an infinite series $\sum_{n\ge 1}(2n-1)!!/[
(2n)!!(7n-2)]$. Since $\sum_{n\ge 1}(2n-1)!!/[(2n)!!(2n+2)]
=1/2$, the integral is smaller than $1/2$. This proves that
the potential \poten\ is negative, as it should be if it
represents the interaction between a vortex and an anti-vortex.
This also proves our earlier claim, that a throat between the
D2-brane and the anti-D2-brane develops in order to minimize
the energy cost.

Another interesting exercise is to compute the correlation 
function $\langle \phi (T)\phi^+(0)\rangle$. We explained
in the previous section that this correlation should vanish.
In Maldacena's conjecture, the corresponding process in the
bulk is a D2-brane moving toward the source D2-branes in
a direction in $R^7$ that is orthogonal to its world-volume,
and touching the horizon at time $X^0=0$, stuck there until
time $X^0=T$, and leaving the horizon. We show that it will
take an infinite amount of time for the test D2-brane to ever
reach the horizon, therefore the above process is impossible.

Take, say $X^5$ as the transverse coordinate in which the test
brane is separated from the source branes. It is a function of
time. Now the induced metric components are $h_{00}=-U^{5/2}/R^2
+R^2U^{-5/2}(\dot{X}^5)^2$, $h_{ij}=R^2U^{-5/2}\delta_{ij}$,
where the spatial coordinates on the test brane are $X^{3,4}$.
Hence $U^2=(X^3)^2+(X^4)^2+(X^5)^2$. The action reads
\eqn\mac{S=-{1\over 4\pi^2g_{YM}^2}\int d^3x\left(1-R^4U^{-5}(\dot{X}^5)^2
\right)^{{1\over 2}}.}
The equation of motion derived from the above action can be solved.
The simplest situation is the point $X^{3,4}=0$. The solution is
\eqn\solu{(X^5)^{{3\over 2}}(t)=a {R^2\over (t-t_0)},}
where $a$ is an integration constant. Thus for this point to reach 
$X^5=0$, it takes an infinite amount of time. Notice that this point
can reach $X^5=\infty$ in finite time.

\newsec{Vortex lines on D3-branes}

As argued in sect.2, a D3-brane intersecting a stack of D3-branes along
a line is a BPS state, the line can be regarded as a vortex line in
the world-volume theory of multi-D3-branes. It is a candidate for a 
trapped 't Hooft loop. We shall compute the static potential between
a vortex line and an anti-vortex line in this section, again utilizing
Maldacena's conjecture.

As in the previous section, set $\ap =1$. The near horizon metric is
\eqn\dthreem{ds^2={U^2\over R^2}ds_4^2+{R^2\over U^2}ds_6^2,}
where $R^2=\sqrt{4\pi gN}$, and the Yang-Mills coupling $g_{YM}^2\sim
g$. The dilaton field is a constant.
The metric is written as though it is a metric on spacetime
$R^4\times R^6$. Due to the $U$ dependent factors, the spacetime
is actually $AdS_5\times S^5$. The supersymmetric gauge theory on N
D3-branes is superconformally invariant, and the anti-de Sitter space
possesses this symmetry explicitly.

The Nambu-Born-Infeld action for a test D3-brane is 
\eqn\dthreea{S=-{1\over (2\pi)^3g}\int d^4x[\det (\p_\alpha X^\mu
\p_\beta X^\nu G_{\mu\nu})]^{{1\over 2}}.}
If the D3-brane is not bent in the two transverse directions $z, \bar{z}$
(which are orthogonal to the intersection line), then it is readily 
checked that the action is identical to that in a flat space.

For a pair of D3-brane and anti-D3-brane separated by a distance $L$
in, say, $x=\Re z$, again we expect that a throat as in fig.2 will
develop. The integration in the action along the intersection line
is trivial, and gives rise to a factor $L_3$ as an infrared cut-off.
We use the same coordinates as in the previous section to parametrize
the remaining two dimensions of the world-volume of this configuration.
The static energy is given by
\eqn\dthreee{E={L_3\over (2\pi)^2gR^2}\int dx U^3\left(1+{R^4\over U^4}
(U')^2\right)^{{1\over 2}}.}

The rest of calculation goes in a similar way as in the previous section.
The minimal $U$ is
\eqn\dthrees{U_0={2R^2\over L}{\Gamma (2/3)\Gamma (1/2)\over
\Gamma (1/6)}.}
And the static potential obtained by subtracting the bare energy from
\dthreee\ is
\eqn\dthreep{V=-{L_3U_0^2\over 2\pi^2 g}\left({1\over 2}-\int_1^\infty
dx x[x^3(x^6-1)^{-{1\over 2}}-1]\right).}
This potential is negative, since it can be shown that the integral
in \dthreep\ is smaller than $1/2$. Now, since $U_0^2\sim R^4/L^2
\sim gN/L^2$, thus the potential scales as $N/L^2$. The dependence
on $g$ drops out, as it should for a D3-brane. 

The power law $L^{-2}$
is simply a consequence of conformal invariance. This says that the
static potential per unit length along the vortex line is proportional
to $L^{-2}$. If we assume that the potential is caused by a point-wise 
potential between a point on the vortex line and a point on the anti-vortex
line, then this point-wise potential must scale as $L^{-3}$.
This is a much softer potential compared with the Coulomb potential
$L^{-1}$.
 
It is straightforward to generalize our calculation to the case of
a finite temperature, where the relevant metric is that of the anti-de
Sitter black hole, as in \refs{\witten, \rty, \bisy}. However, the 
integrals involved in this calculation
can not be explicitly carried out. To check whether the potential is
negative one has to perform some numerical calculations.

\newsec{Vortices on M-branes}

Supersymmetry preserved by a M2-brane lying along the plane
$(X^1,X^2)$ is subject to constraint $\epsilon =\gamma^{012}
\epsilon$, where $\epsilon$ is a Majorana spinor with 32 components.
Just as two intersecting D2-branes, two intersecting M2-branes
preserve $1/4$ of SUSY. For N coincident M2-branes, the equation
$zw^N=c$ again is a solution to the Nambu action. Because of
the branch-cut, we need to perform a singular gauge transformation.
The world-volume theory is believed to be a nonabelian conformal 
field theory, and can be obtained from the D2-brane theory by
taking the strong coupling limit.

We shall again compute the static potential between a vortex and
an anti-vortex, using the anti-de Sitter background. The fundamental
scale is the Planck length $l_p$. It drops out in our calculations
so we set it to equal to $1$. The near horizon metric of the 
N M2-branes is
\eqn\mtwom{ds^2={r^4\over R^4}ds^2_3+{R^2\over r^2}ds_8^2,}
where $r$ is the radial coordinate on $R^8$ which is transverse
to the world-volume of the source branes. $R=(2^5\pi^2 N)^{1/6}$.
Although the above metric is written in a form as though the topology
of spacetime is $R^3\times R^8$, due to the nontrivial $r$
dependent factors in the metric, the spacetime is actually
$AdS_4\times S^7$.

The Nambu action for a test M2-brane is
\eqn\mtwoa{S=-c_2\int d^3x[\det (\p_\alpha X^\mu\p_\beta X^\nu
G_{\mu\nu})]^{{1\over 2}},}
where $c_2$ is a pure number appearing in the M2-brane tension
formula $T^M_2=c_2l_p^{-3}$. Since the M2-brane always has
extension in the $r$ direction, it does not couple to the 
nonvanishing three-form field.

Again we use $\theta$ and $x=\Re z$ to parametrize the world-volume
of the smoothly connected M2-brane and anti-M2 brane, as in
fig.2. The static energy of this configuration is
\eqn\mtwoe{E=2\pi c_2\int dx{r^4\over R^3}\left(1+{R^6\over r^6}
(r')^2\right)^{{1\over 2}}.}
Solving the equation of motion, we obtain the minimal $r_0$, the 
size of the throat, 
\eqn\mtwos{r^2_0={R^3\over L}{\Gamma (3/4)\Gamma (1/2)\over 
\Gamma (1/4)}.}
The static potential obtained by subtracting the bare energy
from \mtwoe\ is
\eqn\mtwopo{V=-4\pi c_2r_0^2\left( {1\over 2} -\int_1^\infty
dx x[x^4(x^8-1)^{-{1\over 2}}-1]\right).}
The integral is equal to $\sum_{n\ge 1}(2n-1)!!/[(2n)!!(8n-2)]$
which is smaller than $1/2$, hence the above potential is
negative. The potential is proportional to $r_0^2$ which in
turn scales as $\sqrt{N}/L$. The behavior $L$ is compatible
with conformal invariance. Just like the D3-brane case, only
$N$ figures into the formula.

Our last example is intersecting M5-branes. A M5-brane intersecting
with a stack of M5-branes along three spatial dimensions preserves
$1/4$ of SUSY. The intersection has codimension $2$ in the
world-volume of coincident M5-branes. This situation is similar
to, but different from the one considered in \ewm, since the 
holomorphic embedding is into a four dimensional space with a trivial
topology. The near horizon geometry induced by the source M5-branes
is
\eqn\mfivem{\eqalign{ds^2&={U^2\over R}ds_6^2+{R^2\over r^2}ds_5^2,
\cr
r&=U^2,\qquad R=(\pi N)^{{1\over 3}}.}}
$r$ is the radial coordinate on $R^5$, the space transverse to 
M5-branes. The spacetime is $AdS_7\times S^4$. We have set $l_p=1$.

An orthogonal M5-brane has three spatial dimensions in common
with the source M5-branes, so does the orthogonal anti-M5-brane.
The integral part of the static energy in these three dimensions
is trivial and gives rise to a factor $V_3$. The nontrivial
part of the energy is
\eqn\mfivee{E=2\pi V_3c_5\int dx{U^5\over R^{3/2}}\left(1+
{4R^3\over U^4}(U')^2\right)^{{1\over 2}}.}

The size of the throat is
\eqn\mfives{U_0={4R^{3/2}\over L}{\Gamma (3/5)\Gamma (1/2)\over
\Gamma (1/10)}.}
The bare energy of the M5-brane and the anti-M5-brane is
$E_0=2V_3c_5\int 2U^3dUd\theta$. This is to be subtracted from
the energy in \mfivee. We find the interaction potential
\eqn\mfivep{\eqalign{V &=-8\pi V_3c_5U_0^4\left({1\over 4}-\int_1^\infty
dxx^3[x^5(x^{10}-1)^{-{1\over 2}}-1]\right)\cr
&=-2\pi V_3c_5U_0^4{\Gamma(3/5)\Gamma(1/2)\over \Gamma(1/10)} .}}
It is negative. The potential per unit three-volume is proportional
to $U_0^4\sim R^6/L^4\sim N^2/L^4$. The power $L^{-4}$ is in
accord with conformal invariance. The point-wise potential
between points on the intersection volumes scales as $L^{-7}$.

Finally, we make the following observation. In all the conformally
invariant cases, the potential is independent of either the coupling
or the Planck length. In the M2, D3, M5 cases, the potential is
proportional to $N^{1/2}$, $N$, $N^2$ respectively. Since the potential
is averaged over colors, the unaveraged result will be $N^{3/2}$, $N^2$
and $N^3$ respectively. These numbers are just the numbers of degrees
of freedom on M2, D3, M5 branes.

\noindent{\bf Acknowledgments} 
We would like to thank D. Minic for a question leading to the
investigation presented here. We are also grateful to E. Martinec 
and V. Sahakian for conversations, and to Yoneya for an exchange
of email.
This work was supported by DOE grant DE-FG02-90ER-40560 and NSF 
grant PHY 91-23780.

\listrefs
\end